\documentclass[12pt]{article}
\usepackage{amsmath}
\usepackage{graphicx,psfrag,epsf,bm}
\usepackage{enumerate}
\usepackage{bm}
\usepackage{amssymb,graphicx,url}
\usepackage[colorlinks,citecolor=blue,urlcolor=blue]{hyperref}
\usepackage{bm}
\usepackage{lipsum}
\usepackage{color}
\usepackage{natbib}
\usepackage{amsmath,here,amsfonts,latexsym,graphicx,here}
\usepackage{amsthm,curves}
\usepackage{graphicx}
\usepackage{caption}
\usepackage{subcaption}
\usepackage{float}
\usepackage{longtable}

\def\l{\lambda}

\def\f{\phi}
\def\d{\delta}
\def\dd{\Delta}

\def\th{\theta}
\def\s{\sigma}

\def\P{{\mathbb P}}

\def\IK{\mathbb K}

\def\bmy{\bm y}

\newtheorem{theorem}{Theorem}

\makeatletter
\def\th@remark{%
	\thm@headfont{\itshape}%
	\normalfont 
	\thm@preskip\topsep \divide\thm@preskip\tw@
	\thm@postskip\thm@preskip
}
\makeatother

\title{\textit{incubation time distribution}}
\usepackage{fancyhdr}
\makeatletter
\let\runtitle\@title
\makeatother
\begin{document}

	\def\spacingset#1{\renewcommand{\baselinestretch}%
		{#1}\small\normalsize} \spacingset{1}
	
	\clearpage\thispagestyle{empty}
	\noindent
	\vspace{0.2cm}
	\begin{center}	
		{\LARGE {\bf Estimation of the incubation time\vspace{0.5cm}
		 distribution for COVID-19}}
	\end{center}
	\vspace*{0.75 cm} 
	\begin{center}
		{\Large Piet Groeneboom}
		\\
		\vspace*{0.25 cm} 
		Delft Institute of Applied Mathematics, Delft University of Technology

	\end{center}
	
\begin{abstract}
We consider smooth nonparametric estimation of the incubation time distribution of COVID-19, in connection with the investigation of researchers from the National Institute for Public Health and the Environment (Dutch: RIVM) of 88 travelers from Wuhan:  \cite{backer:20}. The advantages of the smooth nonparametric approach  w.r.t.\ the parametric approach, using three parametric distributions (Weibull, log-normal and gamma) in \cite{backer:20} is discussed.

It is shown that the typical rate of convergence of the smooth estimate of the density is $n^{2/7}$ in a continuous version of the model, where $n$ is the sample size. The (non-smoothed) nonparametric maximum likelihood estimator (MLE) itself is computed by the iterative convex minorant algorithm (\cite{piet_geurt:14}). All computations are available as {\tt R} scripts in \cite{github:20}.
\end{abstract} 
	
\vspace*{1 cm}
\noindent%
	
\noindent%
{\it  Running headline: incubation time distribution}
	
\newpage

\section{Introduction}
\label{section:Introduction}
Researchers from the Centre for Infectious Disease Control and Prevention of the National Institute for Public Health and the Environment (Dutch: RIVM) analyze in \cite{backer:20} a data set of $88$ travelers who are assumed to have picked up the COVID-19 virus in Wuhan. The distribution of their incubation times is estimated using certain simple distributions, like Weibull, log-normal and gamma.
If the only thing we know about the start of the incubation time is that it belongs to an interval $[0,E_i]$, the log likelihood for one observation is:
\begin{align*}
\log \int_{t\in[0,E_i]}g(S_i-t)\,dF_i(t).
\end{align*}
Here $E_i$ would be the upper bound for the exposure interval, for which we take (looking back)  $0$ as the left point for the $i$th individual (see \cite{tom_gianpi:19}), $S_i$ is the time where the person becomes symptomatic (note that both $S_i\le E_i$ and $S_i>E_i$ can occur), and $F_i$ would be the distribution function of the time of a possible contact with an infector. The exit times and times of becoming symptomatic of the $88$ Wuhan travelers are shown in Table \ref{tab:Wuhan_travelers}.

\begin{longtable}{|c|c|c|c|c|c|c|}
\hline
$i$  & $E_i$ & $S_i$ & $\qquad\qquad\qquad$ & $i$  & $E_i$  & $S_i$\\
\hline
    1  &       5  &       5    & &       45  &     39  &    40  \\ 
    2  &      30  &      33    & &       46  &     35  &    42  \\ 
    3  &      21  &      22    & &       47  &      2  &     6  \\ 
    4  &       1  &       4    & &       48  &     36  &    37  \\ 
    5  &       1  &       6    & &       49  &     38  &    39  \\ 
    6  &       8  &       8    & &       50  &      1  &     8  \\ 
    7  &       4  &       4    & &       51  &     38  &    41  \\ 
    8  &       3  &       3    & &       52  &     38  &    41  \\ 
    9  &      33  &      34    & &       53  &     38  &    39  \\ 
   10  &      33  &      34    & &       54  &     11  &    11  \\ 
   11  &       8  &       8    & &       55  &     36  &    39  \\ 
   12  &       1  &       4    & &       56  &     11  &    11  \\ 
   13  &      20  &      21    & &       57  &     40  &    41  \\ 
   14  &      20  &      28    & &       58  &     36  &    37  \\ 
   15  &      30  &      32    & &       59  &     36  &    41  \\ 
   16  &      35  &      38    & &       60  &     36  &    39  \\ 
   17  &       3  &       7    & &       61  &     27  &    31  \\ 
   18  &      35  &      37    & &       62  &     38  &    40  \\ 
   19  &      36  &      38    & &       63  &     36  &    42  \\ 
   20  &      31  &      38    & &       64  &     40  &    43  \\ 
   21  &      34  &      35    & &       65  &     41  &    43  \\ 
   22  &      29  &      31    & &       66  &     37  &    43  \\ 
   23  &      36  &      37    & &       67  &      1  &     7  \\ 
   24  &       3  &       8    & &       68  &     40  &    42  \\ 
   25  &       7  &       9    & &       69  &     40  &    42  \\ 
   26  &      38  &      39    & &       70  &     31  &    39  \\ 
   27  &      30  &      36    & &       71  &     40  &    41  \\ 
   28  &      28  &      36    & &       72  &     40  &    41  \\ 
   29  &      35  &      36    & &       73  &     41  &    42  \\ 
   30  &      33  &      34    & &       74  &     41  &    43  \\ 
   31  &       3  &       8    & &       75  &      4  &     5  \\ 
   32  &       2  &       4    & &       76  &      4  &     5  \\ 
   33  &       2  &       5    & &       77  &     40  &    41  \\ 
   34  &       5  &       5    & &       78  &     36  &    40  \\ 
   35  &      36  &      37    & &       79  &     36  &    40  \\ 
   36  &      31  &      35    & &       80  &     40  &    42  \\ 
   37  &      41  &      42    & &       81  &     36  &    42  \\ 
   38  &      41  &      42    & &       82  &     38  &    43  \\ 
   39  &       3  &       4    & &       83  &      2  &     9  \\ 
   40  &      38  &      39    & &       84  &     38  &    43  \\ 
   41  &      39  &      41    & &       85  &     37  &    43  \\ 
   42  &      39  &      41    & &       86  &     41  &    42  \\ 
   43  &      39  &      41    & &       87  &     40  &    43  \\ 
   44  &      33  &      39    & &       88  &     40  &    43  \\
\hline
\caption{Exit times and times of becoming symptomatic of the $88$ Wuhan travelers after shifting the entrance
times to $0$.}
\label{tab:Wuhan_travelers}
\end{longtable}

It is clear that, without further assumptions, $g$ and $F_i$ are not identifiable. To remedy this, we assume, as in \cite{backer:20} (see also \cite{reich:09}), that $F_i$ is the uniform distribution on $[0,E_i]$. If we want to use maximum likelihood,  we have to maximize
\begin{align*}
\sum_{i=1}^n\log\left\{\int_{t=0}^{E_i}g(S_i-t)\,dt/E_i\right\},
\end{align*}
and since the $E_i$ do not matter in the maximization problem, we end up with the problem of maximizing
\begin{align}
\label{loglikehood}
\sum_{i=1}^n\log\left\{\int_{t=0}^{E_i}g(S_i-t)\,dt\right\}
\end{align}
where $g$ is the density of the incubation time.

So we deal with the following model. We have an exit time $E_i$ for the exposure interval, an infection time $V_i$ and an incubation time $W_i$. The time of becoming symptomatic is denoted by $S_i$, and $S_i$ is assumed to be the independent sum of $V_i$ and $W_i$.
Our observations are
\begin{align}
\label{def1}
\left(E_i,S_i,\dd_i\right),\qquad i=1,\dots,n,
\end{align}
where $n$ is the sample size and where the indicator $\dd_i$ is defined by
\begin{align}
\label{de21}
\dd_i=1_{\{S_i\le E_i\}},\qquad i=1,\dots,n.
\end{align}
 Using the present notation, the log likelihood for the incubation time distribution function $G$ becomes
\begin{align}
\label{loglikehood2}
\ell(G)=\sum_{i=1}^n\left[\dd_i\log G(S_i)+(1-\dd_i)\log\left\{G(S_i)-G(S_i-E_i)\right\}\right].
\end{align}
Note that the time of becoming symptomatic is still in Wuhan if $\dd_i=1$.

The algorithms we used for analyzing the data set can be found on \cite{github:20}. We describe the data files given there.
The original data file is  {\tt data\_Wuhan\_tsv}, which gives details on the persons in the sample and which can be found in  \cite{backer:20}. This was transformed into a data file {\tt transformed\_data\_Wuhan.txt}, consisting of three columns, giving, respectively, the arrivals in (if available) and departures from Wuhan and the time the person became symptomatic. If the arrival time was not available (possibly because the person was a Wuhan resident), this time was set to $-18$, which means $18$ days before December 31, 2019, which is the zero on the time scale.  For traveler number 67, who apparently had a connecting flight, the duration of stay in Wuhan was changed from 0 to 1 day. This, in turn, was transformed into the input file {\tt inputdata\_Wuhan.txt}, where the time, spent in Wuhan, was shifted making the left point equal to zero, and consists of two columns: the first colums contains the data $S_i-E_i$ (time of becoming symptomatic minus exit time from Wuhan) and $S_i$, time of becoming symptomatic, where all times are shifted to have entrance time zero. If the person became symptomatic in Wuhan we put $E_i$ equal to $S_i$, so $S_i-E_i=0$.

Assuming that the distribution of the possible time of infection is uniform on the exposure interval, and estimating the distribution function $G$ by the Weibull distribution, parametrized as
\begin{align}
\label{Weibull}
G(x)=G_{a,b}(x)=1-\exp\left\{-bx^a\right\},\qquad x>0,
\end{align}
we get as our maximum likelihood estimaters of the parameters $a$ and $b$:
\begin{align}
\label{Weibull_par}
\hat a= 3.03514,\qquad \hat b=0.002619.
\end{align}

Using the Weibull maximum likelihood method, the estimate was computed by two methods. One is a very simple method using {\tt Weibull.cpp}, which is used in {\tt analysis\_EM.R} and  {\tt analysis\_ICM.R}, where also the nonparametric estimate to be discussed in the next sections is computed. For this ``pattern search'' algorithm for looking for the parameters of the Weibull distribution one does not have to compute the derivatives of the log likelihood. It is based on the Hooke-Jeeves algorithm. The other one can be found in {\tt R{\textunderscore}Weibull{\textunderscore}Estimation.R}, where we use the {\tt R} package {\tt lbfgs}, and where the gradient (derivatives of the log likelihood) has to be provided.

The results obtained for the Weibull distribution approach of the two algorithms are remarkably similar. The values in (\ref{Weibull_par}) were produced by the {\tt R} script in \cite{github:20}, using the Hooke-Jeeves algorithm. For a convergence proof of the Hooke-Jeeves algorithm and interesting further discussion of the pattern search algorithms, see \cite{torczon2003} and \cite{torczon:97}.

\begin{figure}[!ht]
\centering
\includegraphics[width=0.5\textwidth]{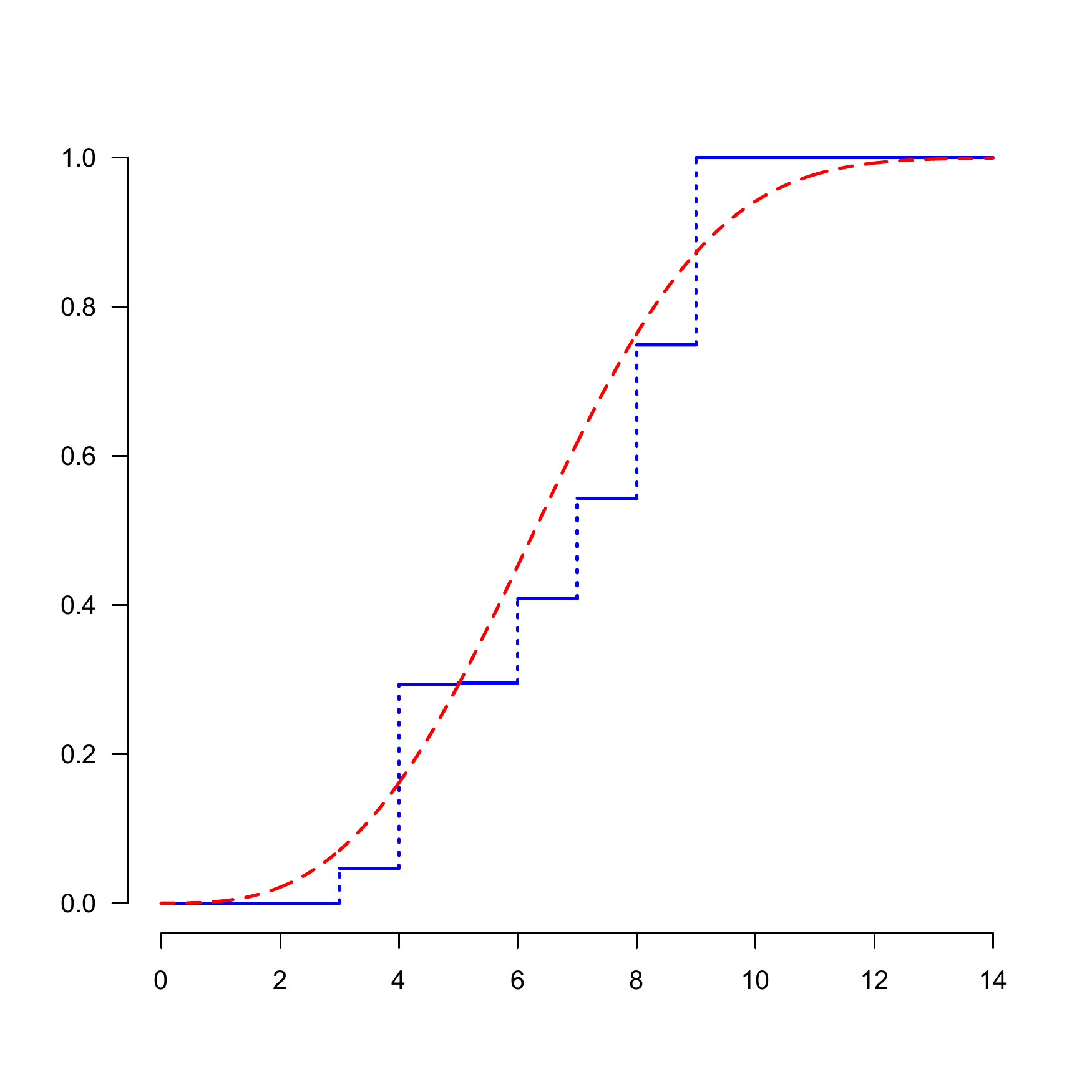}
\caption{The nonparametric maximum likelihood estimate (MLE) $\hat G_n$ of the incubation time distribution function (blue), and the MLE using the Weibull distribution (red, dashed), for the data set analyzed in \cite{backer:20}.}
\label{figure:MLE_Wuhan_data}
\end{figure}

The aim of the present paper, however, is to draw attention to the nonparametric maximum likelihood estimator (the MLE) of the incubation time distribution, which is often also denoted by NPMLE (Nonparametric Maximum Likelihood Estimator). This is the distribution function $\hat G_n$, maximizing (\ref{loglikehood2})  over all distribution functions $G$. The problem of maximizing (\ref{loglikehood2}) over all distribution functions $G$ instead of just Weibull, log-normal or gamma distribution functions is non-trivial and discussed in Section \ref{section:ICM}. We also discuss the smooth estimators based on the MLE, the so-called SMLE (Smoothed Maximum Likelihood Estimator) and the nonparametric density estimator, based on the MLE.

When we want to get an idea of properties of the incubation time distribution, there are (at least) three approaches.

\begin{enumerate}
\item We ``fit'' the data with a parametric distribution from a well-known famiily of distributions like the Weibull, log-normal or gamma distributions. The big disadvantage of this approach is that one usually does not have a good argument for choosing one of these distributions and that important aspects of the data might be completely hidden by the choice of such a distribution.

Convincing examples of this situation are given in Chapter 1 of \cite{silverman:86}.
If one fits the multimodal distribution of the eruptions of the Old Faithful Geyser in Yellowstone Park, Wyoming,  by a unimodal distribution, one will only see one mode instead of the multiple modes that really are there. In that chapter also other interesting examples of how special aspects of the data are revealed by nonparametric density estimation are given.

In fact, estimates of the simple parametric type such as the Weibull, etc.\ will usually be {\it inconsistent}: no matter how many observations one has, there will not be convergence to the right distribution. The ubiquitous appearance of the normal distribution has a completely different origin: the central limit theorem. But this reasoning will generally not apply in the same way for fitting with the Weibull, etc.\ distribution.

Another disadvantage which clearly shows up if people use this method (as in \cite{backer:20}) is that one usually has to introduce several families of distributions (gamma, log-normal, Weibul $\dots$), because there is no compelling reason to pick one of these.

\item We compute the nonparametric MLE. The result for the Wuhan data is shown in Figure \ref{figure:MLE_Wuhan_data} and the bar chart of the point masses of the MLE is shown in Figure \ref{figure:bar_chart_MLE} (the values of the point masses are shown in Table \ref{tab:discrete_MLE}).

This is what one gets if one makes no assumptions at all about the distribution function and this is the ``antipode'' of the fitting with the Weibull etc.\ distribution. Figure \ref{figure:bar_chart_MLE} clearly shows a bimodal discrete density, but one wonders: is this bimodality due to chance fluctuations or is it real? Note that this discrete density is rather different from the density estimation of \cite{silverman:86}, mentioned in point 1. In the latter case one assumes the existence of a (continuous) density with respect to Lebesgue measure instead of a discrete density.

How do we view the distribution of the incubation time? My own inclination is to assume the existence of a continuous density with respect to Lebesgue measure for the incubation time distribution and to use methods as in \cite{silverman:86}, which entails smoothing. Which takes us to:

\item We estimate the density of the incubation time with respect to Lebesgue measure in a nonparametric way.
In this case we also need an extra parameter, the {\it smoothing parameter} or {\it bandwidth}. Now one could argue (as has been done): ``Ah, you objected in point 1 to the use of parametric distributions such as for example the Weibull distribution, but now you introduce a parameter again, the bandwidth!". Fair enough, but: {\it ``The bandwidth is a parameter of a totally different nature than the parameters of the Weibull distribution!''}. With the bandwidth one tries to mediate between the noise and the bias, something we cannot do with the nonparametric estimate, introduced in point 2. Moreover, we can do this in a data-adaptive way, to create independence of a priori assumptions, a type of independence we cannot achieve with the estimates in point 1 above.

We must add, however, that the density estimation problem here is considerably more difficult than the density estimation problems considered in \cite{silverman:86}. This is caused by the fact that our observations are indirect; we assume that the infection took place during the stay in Wuhan, but we do not know when. We only have an interval for this infection time. For this situation we have to use the so-called {\it interval censoring model}, which is for example discussed in \cite{piet_geurt:14}. In fact, we have to deal with a combination of interval censoring (the infection time is contained in an interval, we cannot observe it directly) and {\it deconvolution}, since we have to extract the information from the sum of the infection time and the incubation time. For this reason we get slower rates of convergence of the density estimate: $n^{2/7}$ instead of the usual rate of convergence in density estimation, which is $n^{2/5}$ (see \cite{silverman:86} for the latter rate). An additional complication is that the observations are usually discretized, but we analyze in the sequel both the continuous model just described in Section \ref{section:continuous_model} and the discretized model for which we cannot hope to achieve rate $n^{2/7}$ at each point.

Similar considerations hold for the SMLE, estimating the distribution function. In this case we also need a bandwidth (smaller than the bandwidth for the density estimate) and the rate will be $n^{2/5}$, which is the rate in ordinary density estimation. So in this sense the SMLE is comparable to an ordinary density estimate and the density estimate for the incubation time distribution is comparable to the ordinary estimate of the derivative of a density.
\end{enumerate}

In this paper we focus on the method, described under point 3 above and give algorithms for computing the estimators. {\tt R} scripts for all these methods are given in \cite{github:20}.

It should be noted that the asymptotic distribution of the MLE itself is unknown. In the continuous (not discretized) model it is expected to have the Chernoff limit distribution (location of the maximum of two-sided Brownian motion minus a parabola), but at present this is unknown, as it also is for the related limit distribution of the MLE in the so-called interval censoring, case 2, model (see \cite{piet_geurt:14}).

But we do not need the limit distribution of the MLE itself for deriving the (normal) limit distributions of the SMLE and density estimate, based on the MLE. As an example, we give the derivation for the limit distribution of the density estimate in the simulation model discussed in Section \ref{section:continuous_model} in the appendix (Section \ref{section:appendix}). The fit of the variances, predicted by the asymptotic theory and the variances coming from the simulation study is remarkably good, see Table \ref{tab:variances} and Figure \ref{figure:variances}.

\section{Algorithms for computing the nonparametric maximum likelihood estimator}
\label{section:ICM}
The EM iterations for the MLE maximizing (\ref{loglikehood}), without making this parametric restriction, are in this case given by:
\begin{align}
\label{iteration_EM}
p_j'=p_j n^{-1} \sum_{i=1}^n1_{\{j\in(S_i-E_i,S_i]\}}\Bigm/\sum_{k\in(S_i-E_i,S_i]}p_k\,,
\end{align}
where the ratios are zero if the denominators are zero.
The implementation of this algorithm for the present situation can be found in {\tt analysis\_EM.R} in \cite{github:20}. 

\begin{figure}[!ht]
\centering
\includegraphics[width=0.5\textwidth]{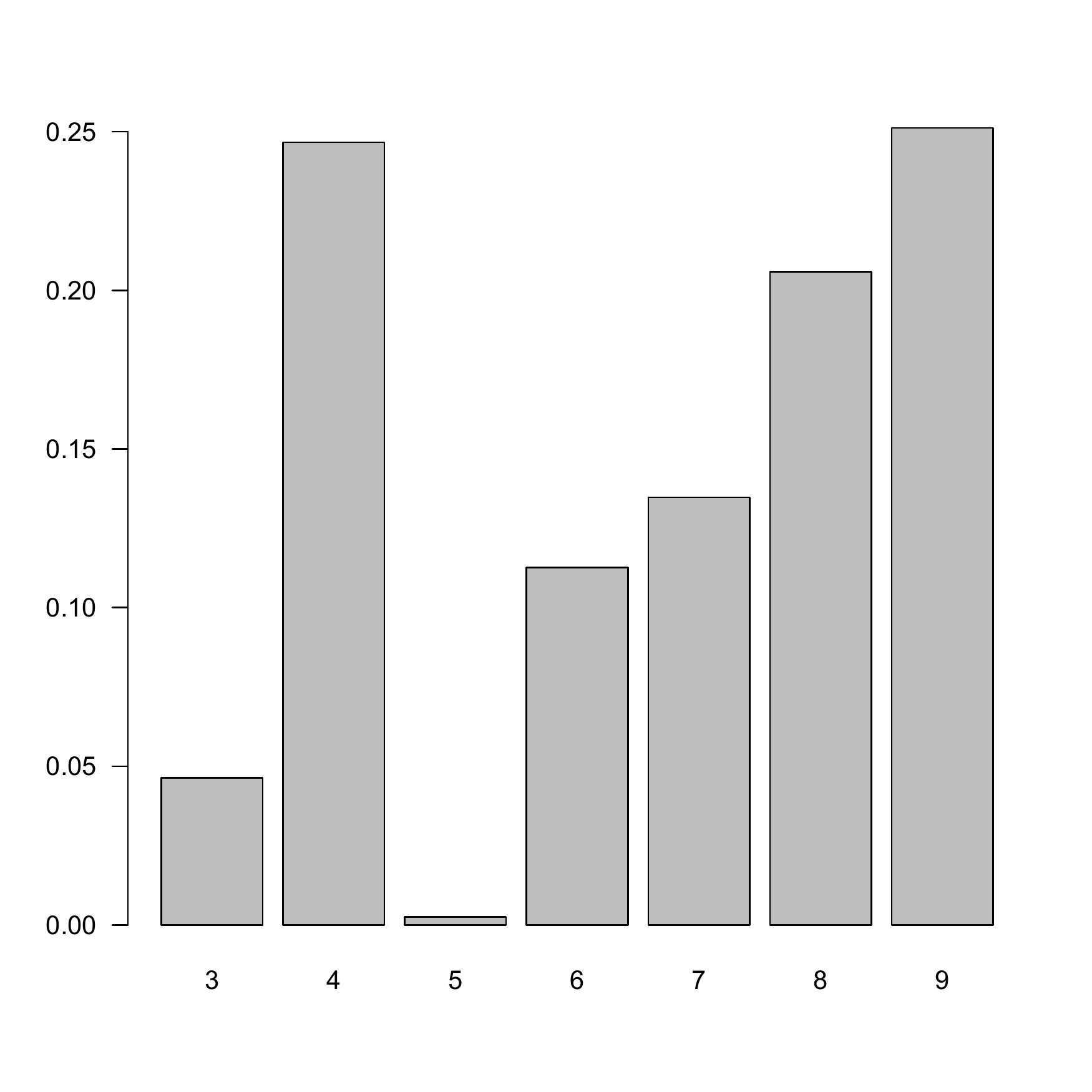}
\caption{Bar chart of the  probability masses of the nonparametric MLE}
\label{figure:bar_chart_MLE}
\end{figure}

The EM iterations were started with the discrete uniform distribution on the $43$ points $1,\dots,43$, which corresponds to the range of values (days) in Table \ref{tab:Wuhan_travelers}, but withdrew its mass after $10,000$ iterations to the $7$ points $3,\dots,9$, which leads to the discrete distribution function, shown in Figure \ref{figure:MLE_Wuhan_data}. A bar chart of the corresponding probability masses is shown in Figure \ref{figure:bar_chart_MLE}. It is seen that this is a bimodal discrete probability distribution with modes at resp. $4$ and $9$ days, with the highest value at the second mode. This discrete probability distribution is also given in Table \ref{tab:discrete_MLE}.

\begin{table}
\begin{center}
\begin{tabular}{|c|c|}
\hline
Number of days  &$p_i$\\
\hline
    3        &0.0463850922\\
    4        &0.2466837048\\
    5        &0.0024858945\\
    6        &0.1126655228\\
    7        &0.1347501680\\
    8        &0.2058210187\\
    9        &0.2512085991\\
\hline
\end{tabular}
\end{center}
\caption{Probability masses of the nonparametric MLE.}
\label{tab:discrete_MLE}
\end{table}

The iteration steps (\ref{iteration_EM}) folllow from the so-called self-consistency equations, which are derived by differentiating the criterion function
\begin{align}
n^{-1}\sum_{i=1}^n \log\left\{\sum_{j\in(S_i-E_i,S_i]}p_j\right\}-\lambda\left\{\sum_{j=1}^{m}p_j-1\right\},
\end{align}
w.r.t. $p_i$, where in this case $m=43$, and $\l$ is a nonnegative Lagrange multiplier, chosen in such a way that
\begin{align}
\label{side_condition}
\sum_{j=1}^{m}p_j=1.
\end{align}
This yields
\begin{align}
\label{EM_conditions}
n^{-1} \sum_{i=1}^n1_{\{j\in(S_i-E_i,S_i]\}}\Bigm/\sum_{k\in(S_i-E_i,S_i]}p_k=\l,\qquad j=1,\dots m,
\end{align}
and multiplying these relations with $p_j$ and summing over $j$ yields $\l=1$, using the side condition (\ref{side_condition}). But the relations (\ref{EM_conditions}) only hold for the {\it active} (in this case $7$) parameters $p_i>0$ of the solution; in the iterations (\ref{iteration_EM}) the inactive parameters $p_i$ will tend to zero.
For more details, see, e.g., \cite{piet_geurt:14}, Section 7.2.

Because of the monotonicity of the distribution function $G$, maximizing the log likelihood over all distribution functions $G$ is an isotonic regression problem, which can be solved by specific isotonic methods. In the present case we can apply the {\it iterative convex minorant algorithm}, discussed in \cite{piet_geurt:14}, Section 7.3.

As discussed in Section \ref{section:Introduction}, the  log likelihood is of type:
\begin{align}
\label{loglikehood1}
f(\bm y)=\sum_{i=1}^m k_i\log\left(G(U_i)-G(T_i)\right),
\end{align}
where $k_i$ is the number of observations $(T_i,U_i)$, and where
\begin{align}
\label{observations_exp}
(T_i,U_i)=\left(0,V_i+W_i\right)1_{\{V_i+W_i\le E_i\}}+\left(V_i+W_i-E_i,V_i+W_i\right)1_{\{V_i+W_i> E_i\}}\qquad i=1,\dots,n,
\end{align}
where $n=88$, and where $V_i$ is the infection time, $W_i$ the incubation time and, as before, $E_i$ the exit time of the travelers from Wuhan, where all observations are centred by subtracting the entrance time.

We first make the so-called preliminary reduction to reduce the problem to a maximization problem in the interior of a convex cone of type
\begin{align*}
\left\{\bmy=(y_1,\dots,y_m)^T:0<y_1\le\dots\le y_m\right\}.
\end{align*}
For the Wuhan data set it can be checked that, without loss of generality, $G(i)=0$, $i\le2$, and $G(i)=1$, $i\ge9$, since in this case values strictly between $0$ and $1$ can only make the likelihood smaller. If we make this preliminary reduction, the log likelihood for the ordered parameters $y_i$, representing the values of the distribution function $G$ at the observation points, becomes:
\begin{align}
\label{loglikehood1a}
f(\bm y)=\sum_{0\le i<j\le7}N_{ij}\log\left(y_j-y_i\right),
\end{align}
where $y_i=G(i+2)$, $i=0,\dots,7$, $y_0=0$, $y_7=1$, and where the triangular array $(N_{ij})$, $0\le i<j\le7$, is given by:
\begin{align*}
\begin{array}{ccccccc}
1   &3   &4   &0   &0   &2   &0 \\  
&2    &1   &0   &0   &0   &9\\   
&&0   &1   &1   &0   &4\\   
&&& 1    &0   &2   &3  \\
&&&&1     &0   &6 \\ 
&&&&& 1     &3   \\
&&&&&& 3 
\end{array}
\end{align*}

We have to maximize (\ref{loglikehood1}) under the restriction $0<y_1\le\dots\le y_6$; by the preliminary reduction, we lost the additional condition $y_6<1$. Let $\bm y=(y_1,\dots,y_6)^T$. The (Fenchel) sufficient and necessary conditions for the solution are:
\begin{align}
\label{fenchel1}
\sum_{j=i}^6 \frac{\partial}{\partial y_j}f(\bm y)\le0,\qquad i=1,\dots,6,
\end{align}
and
\begin{align}
\label{fenchel2}
\sum_{i=1}^6 y_i\frac{\partial}{\partial y_i}f(\bm y)=0,
\end{align}
where $f$ is defined by (\ref{loglikehood1}). 
Since the values $y_i$ are strictly between $0$ and $1$, (\ref{fenchel2}) can only hold if also 
\begin{align*}
\sum_{i=1}^6 \frac{\partial}{\partial y_i}f(\bm y)=0,
\end{align*}
and we can therefore turn (\ref{fenchel1}) into
\begin{align}
\label{fenchel3}
\sum_{j=1}^i \frac{\partial}{\partial y_j}f(\bm y)\ge0,\qquad i=1,\dots,6.
\end{align}
The resulting (nonparametric) MLE $\hat F_n$ is shown in Figure \ref{figure:MLE_Wuhan_data}, together with the MLE assuming that $G$ is a Weibull distribution. The EM algorithm and the iterative convex minorant (ICM) algorithm give exactly the same solutions, but the ICM algorithm needs less iterations ($106$ in this case; the EM algorithm needs between 1000 and $10,000$ iterations).

To compute the MLE via the iterative convex minorant algorithm, we have to construct so-called cusum (cumulative sum) diagrams. The cusum diagram consists of the point $(0,0)$ and the points
\begin{align}
\label{cusum}
\sum_{j=1}^i \left(w_j,\frac{\partial}{\partial y_j}f(\bm y)+w_j y_j\right),\qquad i=1,\dots,6,
\end{align}
where
\begin{align}
\label{weights}
w_j=-\frac{\partial^2}{\partial y_j^2}f(\bm y).\qquad j=1,\dots,6.
\end{align}
At each iteration step the left derivative vector $\bm y'$ of the greatest convex minorant of the cusum diagram is computed on the basis of the current value $\bm y$, and the stationary point of this iteration is the solution of the optimization problem. We perform line search in case the full step to $\bm y'$ would not lead to improvement or would go out of bounds. For more theory, see \cite{piet_geurt:14}. 

\begin{figure}[!ht]
\begin{subfigure}[b]{0.4\textwidth}
\includegraphics[width=\textwidth]{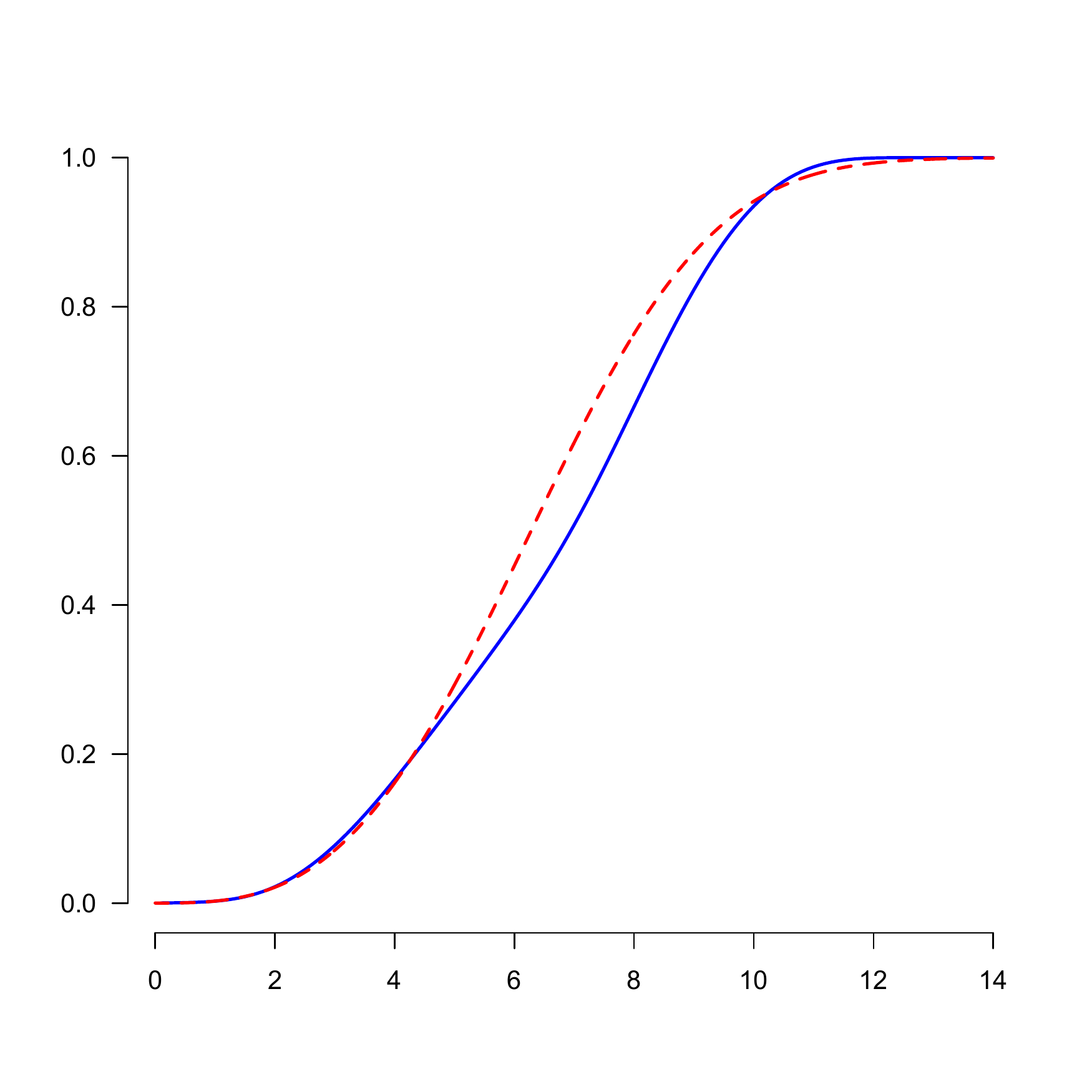}
\caption{}
\label{fig:crossval1}
\end{subfigure}
\hspace{1cm}
\begin{subfigure}[b]{0.4\textwidth}
\includegraphics[width=\textwidth]{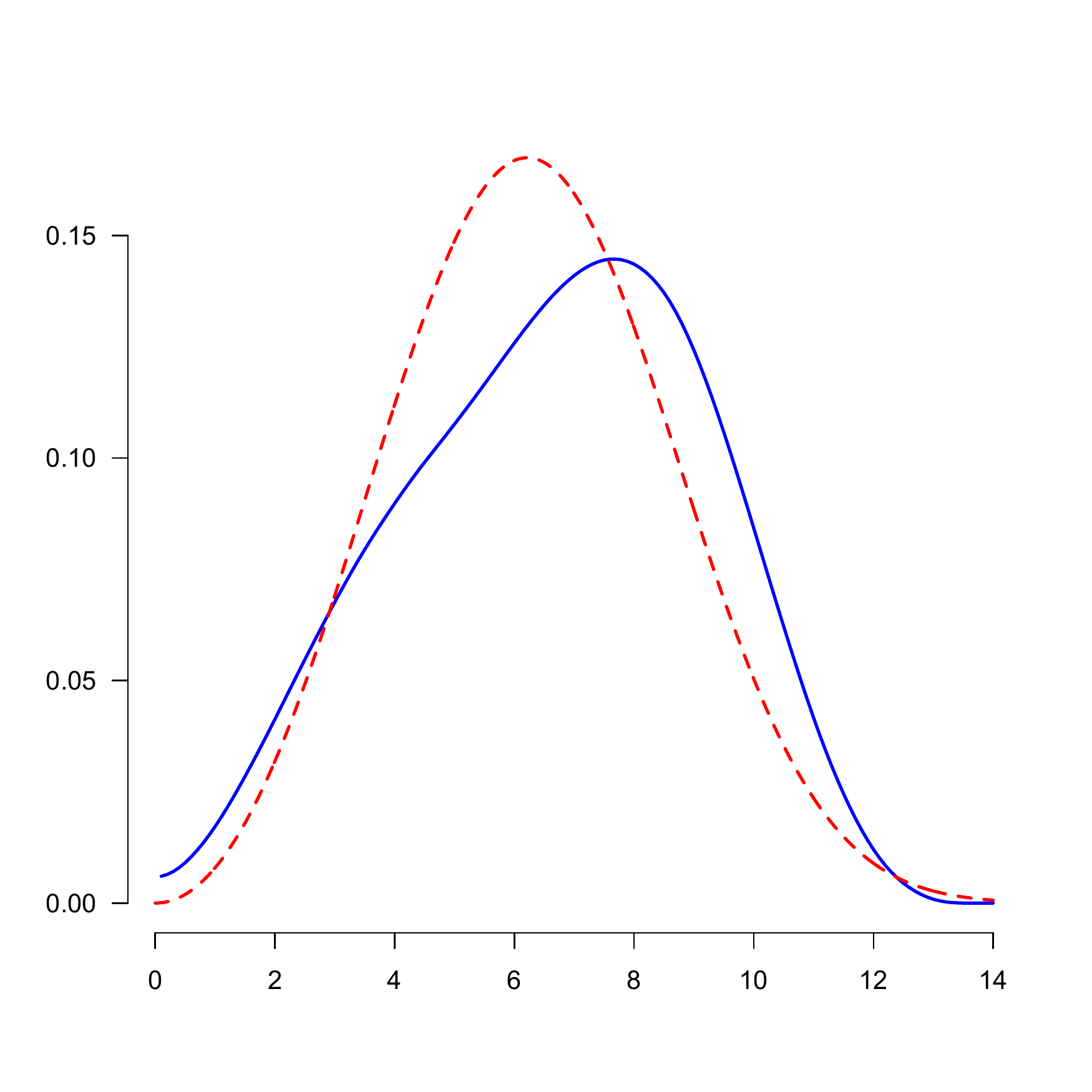}
\caption{}
\label{fig:exit_dens1}
\end{subfigure}
\caption{(a): The smoothed nonparametric maximum likelihood estimate (SMLE) of the incubation time distribution function (blue), and the MLE using the Weibull distribution (red, dashed), for the data set analyzed in \cite{backer:20} and (b): the smoothed nonparametric maximum likelihood estimate of the incubation time density function (blue), and the MLE of the density using the Weibull distribution (red, dashed), for the data set analyzed in \cite{backer:20}.}
\label{figure:SMLE_density}
\end{figure}

As in \cite{piet_geurt:14}, section 1.2, we can compute the smoothed maximum likelihood estimator (SMLE) and also an estimate of the density. The SMLE is defined by
\begin{align}
\label{SMLE}
\tilde G_{nh}(t)=\int \IK((t-y)/h)\,d\hat G_n(y),
\end{align}
where $h>0$ and $\IK$ is an integrated kernel
\begin{equation}
\label{first_int_kernel}
\IK(x)=\int_{-\infty}^x K(u)\,du.
\end{equation}
Here $K$ is a symmetric kernel with support $[-1,1]$, for example the triweight kernel
\begin{align}
\label{triweight}
K(u)=\frac{35}{32}\left(1-u^2\right)^31_{[-1,1]}(u).
\end{align}
We estimate the density by
\begin{align}
\label{dens_MLE}
\tilde g_{nh}(t)=h^{-1}\int K((t-y)/h)\,d\hat G_n(y).
\end{align}
For the present analysis we took $h=3.6$ in (\ref{SMLE}) and $h=4.6$ in (\ref{dens_MLE}); these bandwidths were chosen by a bootstrap method, explained in Section 
\ref{section:bandwidth_choice}. The resulting estimates are shown in Figure \ref{figure:SMLE_density}.

\section{Data-adaptive bandwidth choice for the density estimate and the SMLE}
\label{section:bandwidth_choice}
Let the random variables $E_i$ with values on the integers (``days'') on the interval $[1,43]$ represent the exit times.
Furthermore, let $V_i$ denote the (unknown) infection time, which we take, conditionally on $E_i$,  to be uniform on $[0,E_i]$, and let $W_i$ denote the (again unkown) incubation time.
Our observations are the triples $(E_i,S_i,\dd_i)$, given by (\ref{def1}).

To determine the bandwidth $h$ of our density estimator
\begin{align}
\label{dens_est_incub}
\hat g_{nh}(t)=\int K_h(t-y)\,d\hat G_n(y),
\end{align}
where $\hat G_n$ is the MLE of the distribution function $G$ of the incubation time, we follow a method somewhat similar to the method used in \cite{SenXu2015}.

\begin{figure}[!ht]
\centering
\includegraphics[width=0.5\textwidth]{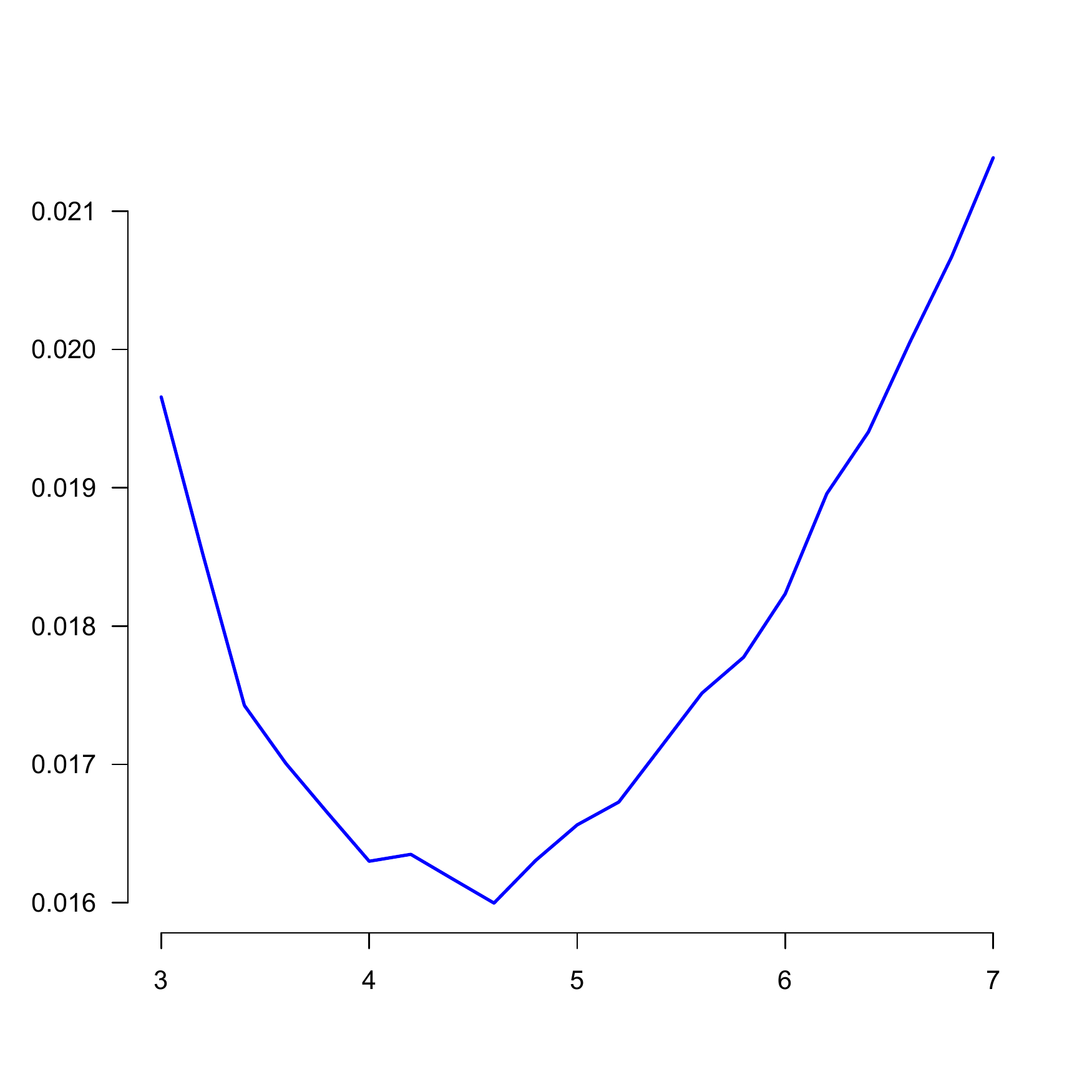}
\caption{ $\hat{\text{MSE}_g}(h)$, given by (\ref{bootstrap_MSE_g}), as function of $h$.}
\label{figure:loss_bootstrap}
\end{figure}

We take $B=10,000$ bootstrap samples of observations $(E_i,S_i^*,\dd_i^*)$, corresponding to the observations $(E_i,S_i,\dd_i)$. The $S_i^*$ are generated as the sums (rounded to the nearest integer) of a Uniform$(0,E_i)$ random variable $V_i^*$ and a random variable $W_i^*$, generated from the density $\hat g_{nh_0}$ by rejection sampling for a fixed $h_0$, for which we took $h_0=4$ in the present case. The $\dd_i^*$ are given by
\begin{align*}
\dd_i^*=1_{\{V_i^*+W_i^* \le E_i\}}\,.
\end{align*}
Note that we keep the $E_i$ the same as in the original sample, somewhat analogously to the procedure followed in \cite{SenXu2015}, which relieves us from the duty to estimate the exit time distribution.

Next we computed
\begin{align}
\label{bootstrap_MSE_g}
\hat{\text{MSE}}_g(h)=B^{-1}\sum_{b=1}^B \int\left\{\hat g^*_{nh}(x)-\hat g_{nh_0}(x)\right\}^2\,dx.
\end{align}
The resulting loss function $\hat{\text{MSE}}_g(h)$ is shown in Figure \ref{figure:loss_bootstrap}, which gave as the minimizing bandwidth $\hat h\approx4.6$. Taking $h_0=3$ in our function of reference $\hat g_{nh_0}$ gave the same minimizing value. The (approximate)  independence of the starting value $h_0$ was also observed for the analogous bandwidth selection procedure in \cite{SenXu2015}.

Similarly, we computed
\begin{align}
\label{bootstrap_MSE_F}
\hat{\text{MSE}}_G(h)=B^{-1}\sum_{b=1}^B \int\left\{\hat G^*_{nh}(x)-\hat G_{nh_0}(x)\right\}^2\,dx,
\end{align}
as a function of $h$ by the same bootstrap procedure, where $\hat G^*_{nh}$ was computed for the bootstrap samples. The integrals were approximated by Riemann sums with step size $0.1$ on the interval $[0,14]$. The {\tt R} scripts for this procedure can again be found on \cite{github:20}.
The method used here is called the ``smoothed bootstrap'', because we generate the bootstrap samples from the smooth estimate $\hat g_{nh_0}$ of the density of the incubation time (added to a uniform$[0,E_i]$ random variable) instead of just resampling with replacement from the data $(E_i,S_i,\dd_i)$, as one would do in the ordinary bootstrap.

A perhaps slightly unorthodox variant of the present method is the smooth bootstrap where we do not round the sums of $V_i^*$ and $W_i^*$ to the nearest integer, but just use them as continuous variables (for more information on the continuous model see the next session). The unorthodox aspect is that, in our bootstrap experiment, we do not recreate exactly the same situation as in our original setting, where the data are integers. In fact, we create data for the continuous model, where we can easier compare bias and variance. We tried this out for the density estimates, and it actually gave exactly the same minimizing bandwidth $h=4.6$ for the least squares criterion. More research on this method is necessary, though.

\section{The continuous model}
\label{section:continuous_model}
Applying the method of the preceding section to the discrete data, where one only uses days on the time axis, is somewhat dubious, since, in fact, we do not have information on a finer scale, which would allow us to let the bandwidth (and therefore the bias) tend to zero. It is conceivable that we have information on a finer scale, for example the time of the outgoing flight or the time of day of becoming symptomatic. Presently both times are interval censored (where one day is the interval). We could therefore introduce another assumption, for example that the time of becoming symptomatic is uniformly distributed over a day.
In any case, there seems to be enough reason to study the continuous model, where one would have (approximately) continuous observations, and to analyze what can be expected in this case.

We define as before the indicator $\dd$ by
\begin{align}
\label{def_indicator}
\dd=1_{\{S\le E\}},
\end{align}
where $E$ is again the exit time and $S$ is the time of becoming symptomatic, and consider the following simulation experiment. $E_i$ is uniform$[0,M]$, the time of infection $V_i$ is a Uniform random variable on $[0,E_i]$, conditionally on $E_i$, and the incubation time $W_i$ is a truncated Weibull$(a,b)$ distribution, where $a$ and $b$ have the same values as the estimates $\hat a$ and $\hat b$ in (\ref{Weibull_par}), respectively, and where the truncation interval $[0,M_1]$ is contained in the interval $[0,M]$. In the present simulation, we took $M_1=20$ and $M=30$. In this way the upper bound for the observations $S_i$ is equal to $50$, which is somewhat comparable with the upper bound $43$ of the observations $S_i$ for the Wuhan travelers. This means that $S_i=V_i+W_i$, where we assume that $V_i$ and $W_i$ are independent, and that our observations are the triples $(E_i,S_i,\dd_i)$.

\begin{figure}[!ht]
\begin{subfigure}[b]{0.4\textwidth}
\includegraphics[width=\textwidth]{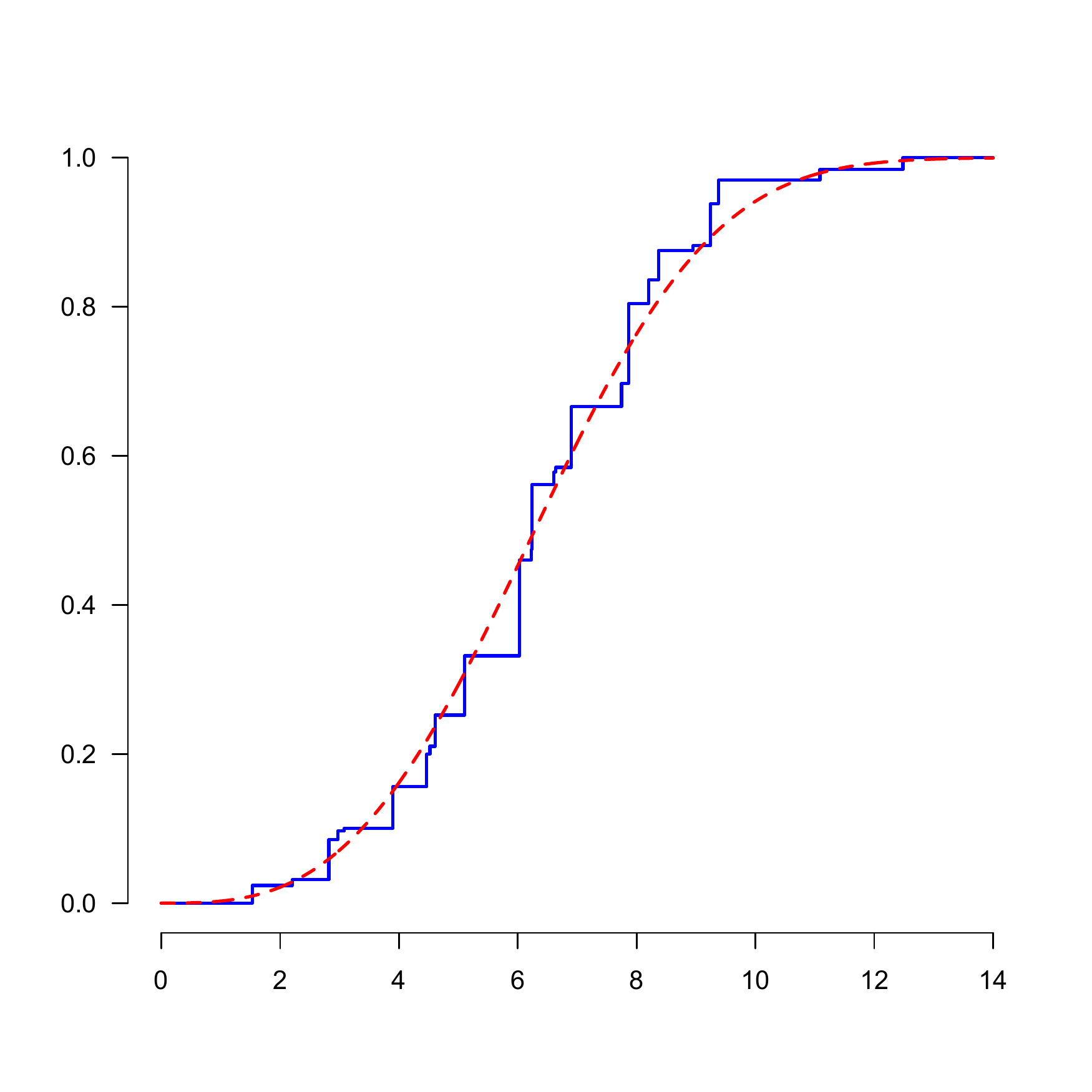}
\caption{}
\label{fig:crossval2}
\end{subfigure}
\hspace{1cm}
\begin{subfigure}[b]{0.4\textwidth}
\includegraphics[width=\textwidth]{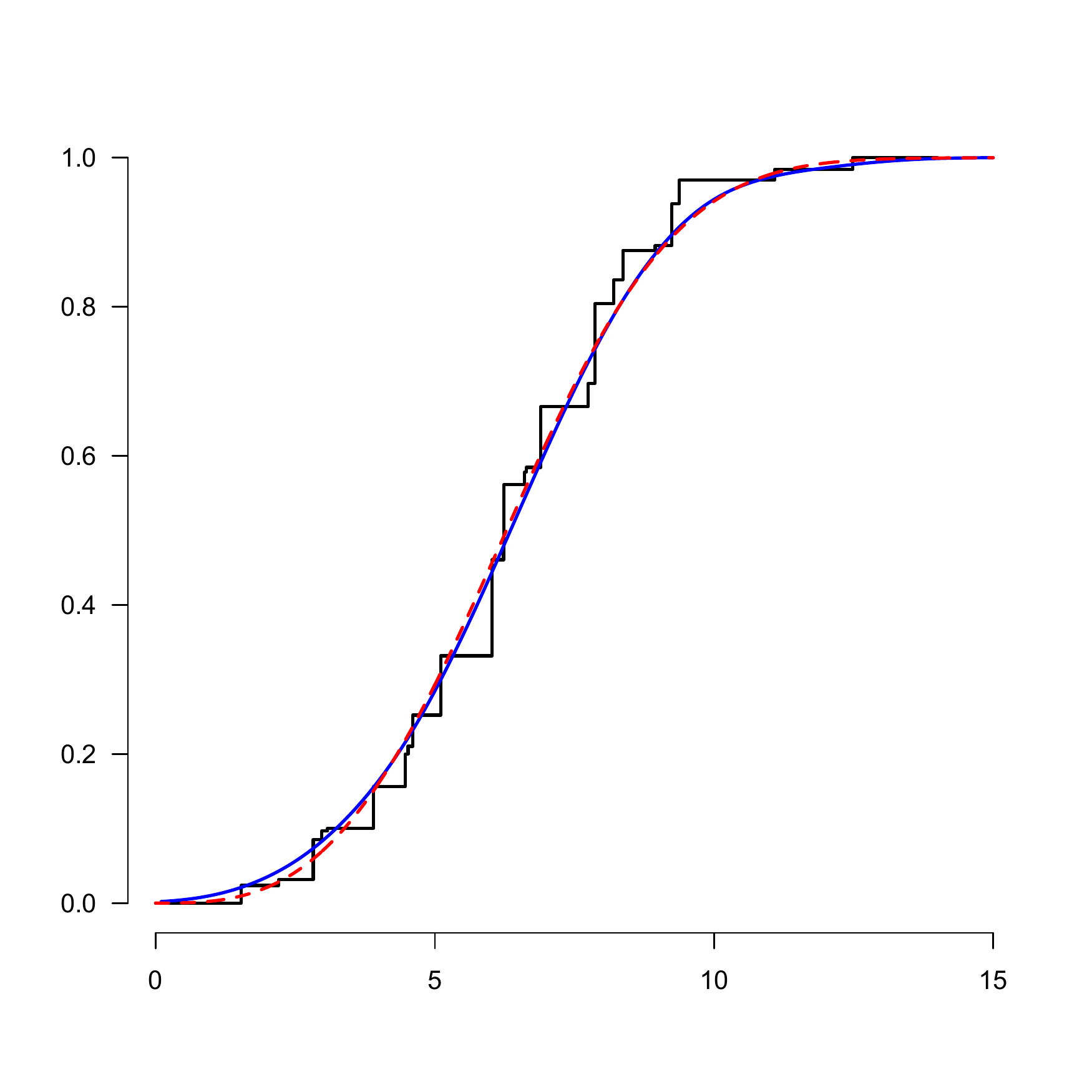}
\caption{}
\label{fig:exit_dens}
\end{subfigure}
\caption{(a): The nonparametric maximum likelihood estimate (MLE) $\hat G_n$ of the incubation time distribution function (blue) for a sample of size $n=1000$, and the truncated Weibull distribution function (red, dashed) with parameters $a$ and $b$, in the simulation model where the variables are not discretized. (b): The MLE (black) and the SMLE (blue), for the same sample, and the truncated (on $[0,M_1]$) Weibull distribution function (red, dashed). The bandwidth of the SMLE is $h=3$.}
\label{fig:continuousMLE1000}
\end{figure}

The MLE of the incubation time, where $E_i$ and $S_i$ are known, looks rather different from the MLE based on the discretized observations shown in Figure \ref{figure:MLE_Wuhan_data}. An example of such an MLE is shown in Figure \ref{fig:continuousMLE1000} for a sample of $n=1000$. Since in this case the MLE can have more jumps, it has the possibility to be much closer to the continuous distribution function. It maximizes again expression (\ref{loglikehood}), but this time the variables $E_i$ and $S_i$ are not discretized.

In this setup, the SMLE will, in the interior of the interval $[0,M_1]$, pointwise have the $n^{2/5}$ rate and the corresponding nonparametric density estimate the $n^{2/7}$ rate of convergence, and the pointwise limit distributions will be normal in both cases (see Section \ref{section:appendix} of the present paper and \cite{piet_geurt:14}, section 11.4). For the density estimate in the present simulation model we get the following result.

\begin{theorem}
\label{theorem1}
Let $\tilde g_{n,h_n}$ be the estimate of the density, defined by
\begin{align*}
\tilde g_{n,h_n}(t)=h^{-1}\int K((t-y)/h_n)\,d\hat G_n(y)=\int K_{h_n}(t-y)\,d\hat G_n(y),
\end{align*}
where $h_n\sim cn^{-1/7}$, for some $c>0$. Let the score function $\th_{t,h,G}$ be defined by
\begin{align}
\label{score1}
\th_{t,h,G}(e,s,\d)=
\d\frac{\f(s)}{G(s)}+(1-\d)\frac{\f(s)-\f(s-e)}{G(s)-G(s-e)\}},
\end{align}
where $\d$ is the indicator $\d=1_{\{s\le e\}}$ and where $\f$ solves the integral equation
\begin{align}
\label{phi-eq2a}
&-\frac{\f(w)}{MG(w)}\log(M/w)
+\frac1{M}\int_{e=0}^w\frac1{e}\left\{\frac{\f(w+e)-\f(w)}{G(w+e)-G(w)}-\frac{\f(w)-\f(w-e)}{G(w)-G(w-e)}\right\}\,de
\nonumber\\
&\qquad\qquad+\frac1{M}\int_{e=w}^M\frac1{e}\frac{\f(w+e)-\f(w)}{G(w+e)-G(w)}\,de\nonumber\\
&=\frac{\partial}{\partial w}K_h(w-t),
\end{align}
defining $0/0=0$. Let $\P_n$ be the empirical probability measure of a sample $(E_1,S_1,\dd_1)$, $\dots$, $(E_n,S_n,\dd_n)$. Then we have, taking $h=h_n\sim cn^{-1/7}$, for a $c>0$,  and $G=G_0$ (the underlying incubation time distribution)  in (\ref{score1}) and (\ref{phi-eq2a}),
\begin{align}
\label{asymptotic_normality1}
n^{2/7}\left\{\tilde g_{n,h_n}(t)-\int K_{h_n}(t-y)\,dG_0(y)\right\}=n^{2/7}\int K_{h_n}(t-y)\,d\bigl(\hat G_n-G_0\bigr)(y)\stackrel{{\cal D}}\longrightarrow N(0,\s^2),
\end{align}
where $N(0,\s^2)$ is a normal distribution with mean zero and variance $\s^2$ given by:
\begin{align*}
\s^2 = \lim_{n\to\infty}\text{var}\left(n^{2/7}\int \th_{t,h_n,G_0}(e,s,\d)\,d\P_n(e,s,\d)\right).
\end{align*}
\end{theorem}
A sketch of the proof is given in the Appendix and the rather good fit of the simulated variance and the variances predicted by this asymptotic result is shown in Table \ref{tab:variances} and Figure \ref{figure:variances}. We do not have an explicit expression for the function $\f$, but could solve the integral equation numerically. In the present simulation study, $G_0$ is given by the truncated Weibull distribution function with parameters given by (\ref{Weibull_par}).

\begin{figure}[!ht]
\centering
\includegraphics[width=0.5\textwidth]{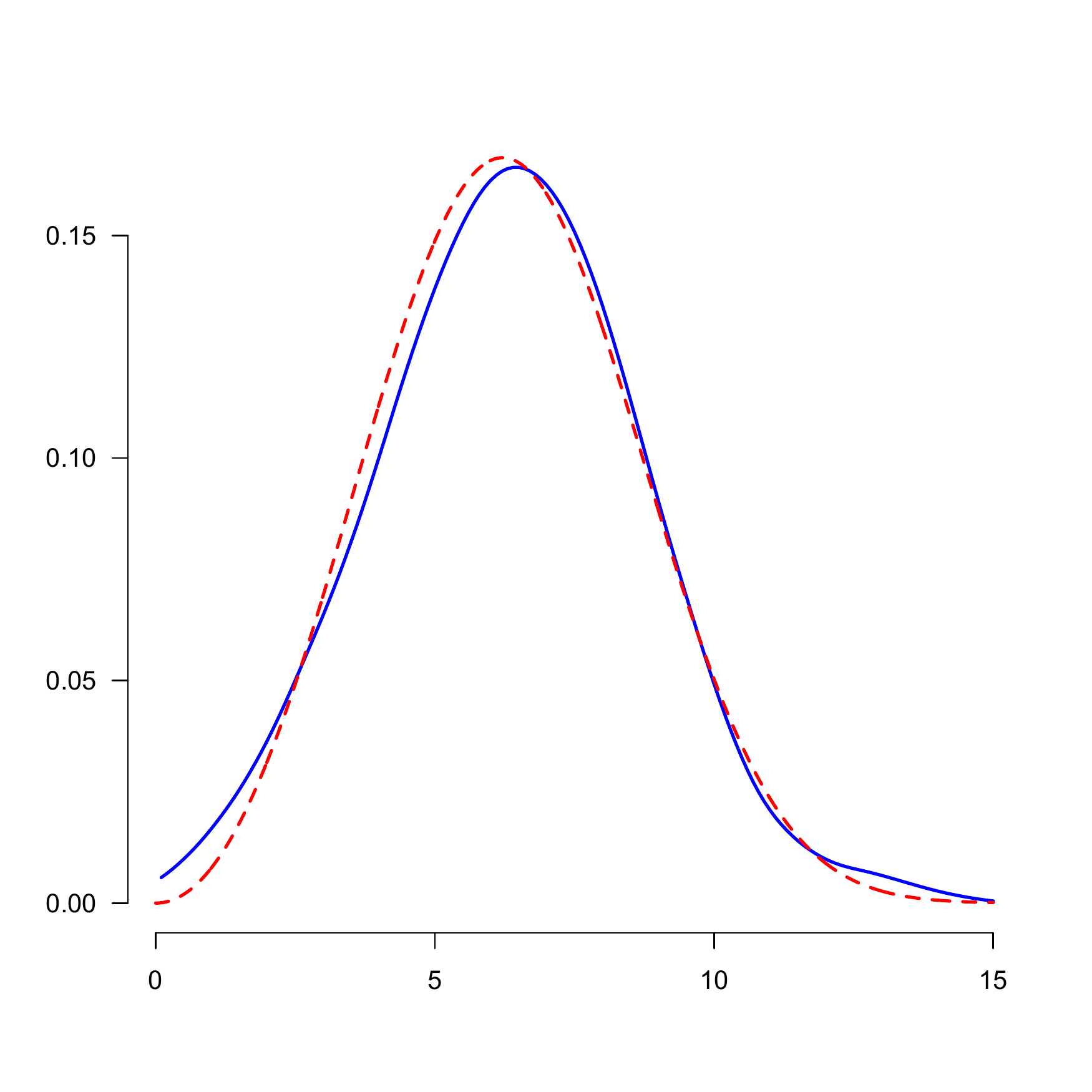}
\caption{The nonparametric estimate of the density of the incubation time (blue, solid), based on a sample of size $n=1000$, based on the truncated Weibull distribution, where we use bandwidth $h=3.4$. The red dashed curve is the truncated Weibull density with parameters $a$ and $b$ of (\ref{Weibull_par}).}
\label{figure:dens1000}
\end{figure}

This means that we can apply the same techniques as in \cite{kim_piet:17EJS} and the {\tt R}-package \cite{curstatCI:17}, and for example compute pointwise bootstrap confidence intervals for the density. The bandwidth was determined by taking bootstrap samples of size $m=50$, using bandwidths of size $cm^{-1/7}$ and using the optimal constant $\hat c$ over the east squares criterion in the bandwidth $\hat c n^{-1/7}=3.51991$, where $n=1000$, for the density in the original sample, where we compare with the density estimate with bandwidth $h=3$ in the original sample. This follows the procedure shown in the vignette of the {\tt R}-package \cite{curstatCI:17}. For the motivation for taking bootstrap samples of a smaller sample size, see \cite{kim_piet:17EJS}. The method goes back to \cite{hall:90}. Since we have a simulation model here, we can also compute the real minimizing $h$, in a comparison with the truncated Weibull density. This yielded $h=3.4$ in the present case, which is a value not far from the bandwidth found by the bootstrap sampling. In the pictures of this section, we took $h=3.4$.

\begin{figure}[!ht]
\centering
\includegraphics[width=0.5\textwidth]{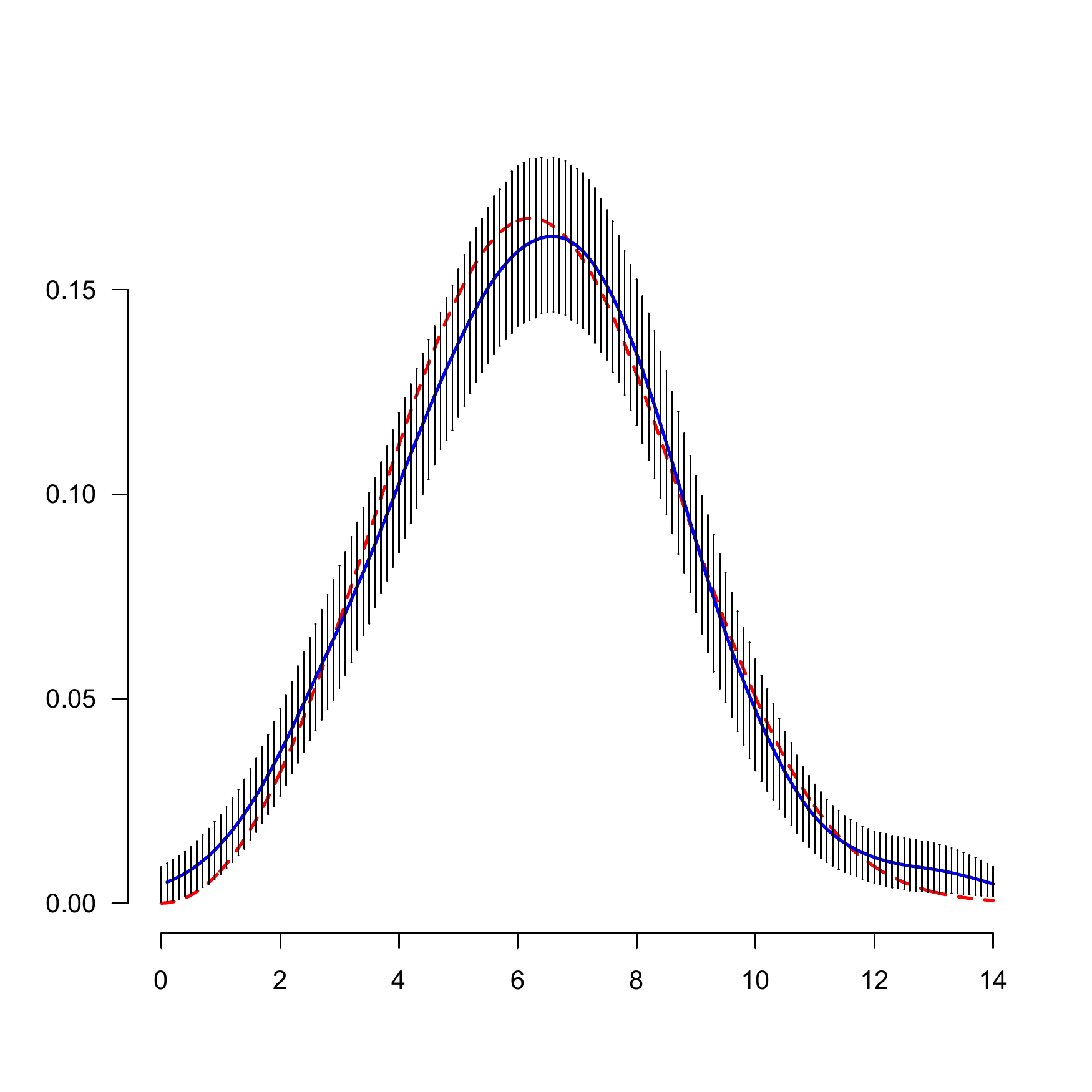}
\caption{Density estimate (blue) and pointwise bootstrap 95\% confidence intervals for the density of the incubation time distribution for a sample of size $n=1000$ (same sample as in Figures \ref{fig:continuousMLE1000} and \ref{figure:dens1000}). The truncated Weibull density is given by the red dashed curve.}
\label{figure:CIdens1000}
\end{figure}

The bootstrap 95\% confidence intervals for the density are shown for a sample of size $n=1000$ in Figure \ref{figure:CIdens1000}. These computations can again be checked on \cite{github:20}. For these  intervals just 1000 bootstrap samples were taken, resampling with replacement from the original sample of triples $(E_i,S_i,\dd_i)$, computing the density estimate again in the bootstrap samples and determining the 2.5\% and 97.5\% percentiles of the values of the density estimates in the $1000$ bootstrap samples. To get really good intervals it is probably necessary to use an asymptotic pivot though, based on Theorem \ref{theorem1}. This matter is subject to further investigation.

\section{Concluding remarks}
\label{section:conclusion}
We offered an alternative nonparametric approach to the estimation of the incubation time distribution which was estimated by parametric methods in \cite{backer:20} for a data set of travelers from Wuhan. In this way we do not have to choose a parametric distribution, like the Weibull, log-normal or gamma, as in \cite{backer:20}, but compute a nonparametric maximum likelihood estimate instead which does not need the arbitrary choice of parameters at all.

However, to give a smooth estimate of the distribution function and (continuous) density, we have to choose a bandwidth parameter. For this choice a smoothed bootstrap approach was suggested. We also considered the model where the observations are not discretized and discussed  rates of convergence, bootstrap confidence intervals and a limit theorem in that case. The present paper can be considered to be the technical companion of the column \cite{pietNAW:20}. All numerical computations are given as {\tt R} scripts in \cite{github:20}.

\section*{Acknowledgements}
\label{section:acknowledgements}
I want to thank Guus Balkema, Ronald Geskus and Siem Heisterkamp and a referee for their comments.

\section{Appendix}
\label{section:appendix}
Using the notation of p.\ 330  of \cite{piet_geurt:14}, we define the score function $\th_{t,h,G}$ by:
\begin{align}
\label{def_theta}
&\th_{t,h,G}(e,s,\d)=E\bigl[a(W)|(E,S,\dd)=(e,s,\d)\bigr]\nonumber\\
&=\d\frac{\int_{w\le s}a(w)\,dG(w)}{G(s)}+(1-\d)\frac{\int_{w\in(s-e,s]}a(w)\,dG(w)}{G(s)-G(s-e)},
\end{align}
where $\d=1_{\{s\le e\}}$. We assume $G(M_1)=1$, where $G$ is the distribution function of the incubation time and $M_1$ is the upper bound of the support of the distribution (taken to be $M_1=20$ in the simulations).

Defining, as in for example the interval censoring model,
\begin{align*}
\f(u)=\int_{y\le u}a(y)\,dG(y),
\end{align*}
we get:
\begin{align}
\label{score}
\th_{t,h,G}(e,s,\d)=
\d\frac{\f(s)}{G(s)}+(1-\d)\frac{\f(s)-\f(s-e)}{G(s)-G(s-e)\}},
\end{align}
where we define $0/0=0$. Note that $\f$ is absolutely continuous w.r.t. $G$ and that $\f(s)=0$, $s\ge M_1$,
since we assume, as usual, $a\in L_2^0(G)$, where $L_2^0(G)$ is the space of square integrable functions $f$ w.r.t. $dG$, with the property $\int f(x)\,dG(x)=0$.

In the present model, the infection time is uniform on $[0,E]$ and $E$ is Uniform$[0,M]$. So we get the following integral equation for the estimation of the density if $w\in[0,M)$, 
\begin{align}
\label{phi-eq}
&E\left[\th_{t,h,G}(E,S,\dd)|W=w\right]\nonumber\\
&=\int_{e\in[w,M]}\frac1{Me}\left\{\int_{s\in[w,e]} \frac{\f(s)}{G(s)}\,ds\right\}\,de
+\int_{e\in[w,M]}\frac1{Me}\left\{\int_{s\in[e,w+e]}\frac{\f(s)-\f(s-e)}{G(s)-G(s-e)}\,ds\right\}\,de
\nonumber\\
&\qquad\qquad
+\int_{e\in(0,w]}\frac1{Me}\left\{\int_{s\in[w,w+e]}\frac{\f(s)-\f(s-e)}{G(s)-G(s-e)}\,ds\right\}
\,de\nonumber\\
&=K_h(w-t).
\end{align}

Differentiation w.r.t.\ $w$ yields for the density estimate:
\begin{align}
\label{phi-eq2}
&-\frac{\f(w)}{MG(w)}\log(M/w)
+\frac1{M}\int_{e=0}^w\frac1{e}\left\{\frac{\f(w+e)-\f(w)}{G(w+e)-G(w)}-\frac{\f(w)-\f(w-e)}{G(w)-G(w-e)}\right\}\,de
\nonumber\\
&\qquad\qquad+\frac1{M}\int_{e=w}^M\frac1{e}\frac{\f(w+e)-\f(w)}{G(w+e)-G(w)}\,de\nonumber\\
&=\frac{\partial}{\partial w}K_h(w-t),
\end{align}

So we get the representation
\begin{align*}
\int K_h(t-y)\,d\bigl(\hat G_n-G_0\bigr)(y)=\int \th_{t,h,\hat G_n}(e,s,\d)\,dP_0(e,s,\d),
\end{align*}
where $\hat G_n$ is the MLE and $\f$ solves (\ref{phi-eq}) for $G=\hat G_n$ (compare to (11.44), p. 331 of \cite{piet_geurt:14}).

\begin{figure}[!ht]
\centering
\includegraphics[width=0.5\textwidth]{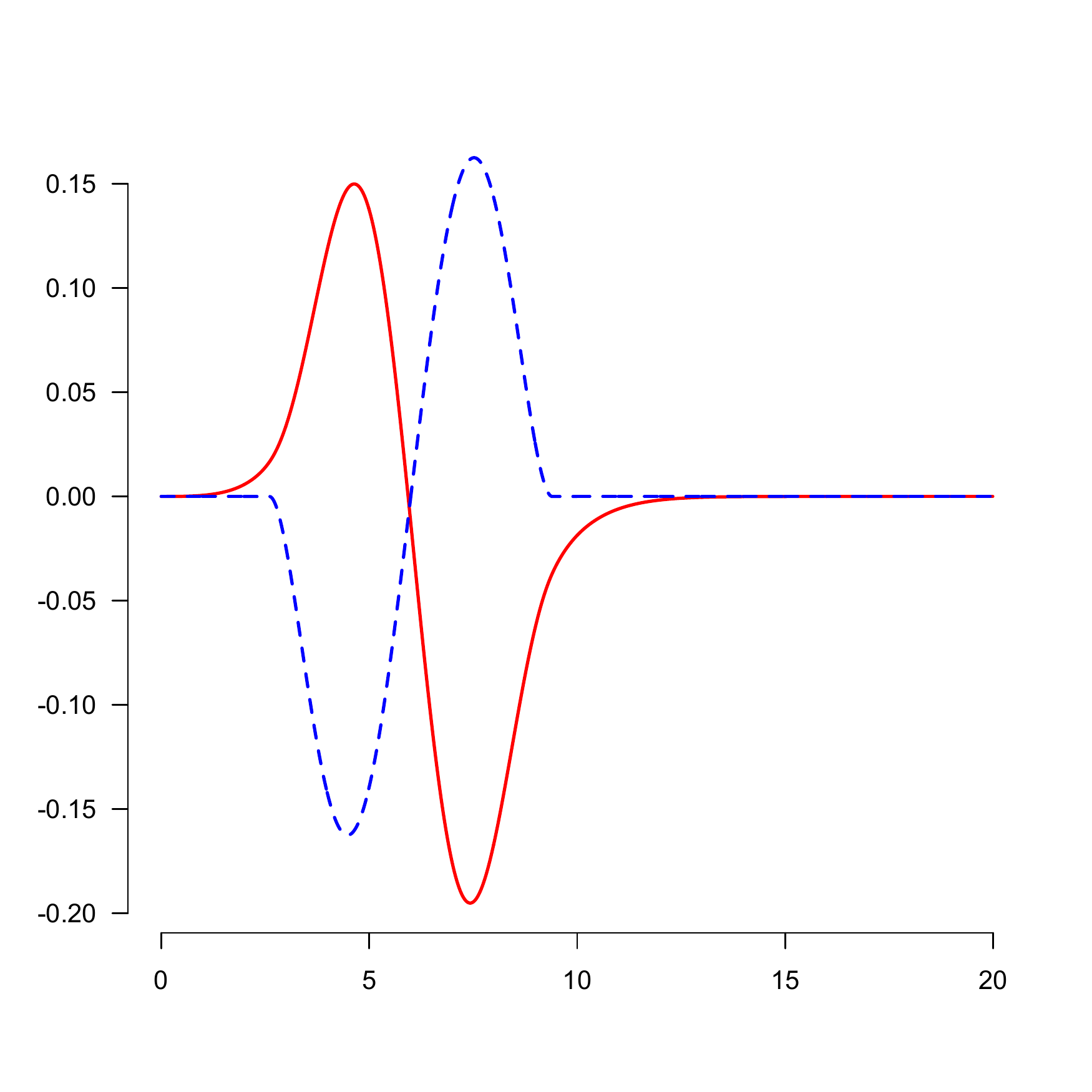}
\caption{The function $\phi$ (red, solid), for $h=3.4$, the triweight kernel $K$, and $t=6$. The blue dashed curve is the function $w\mapsto \frac{\partial}{\partial w}K_h(w-t)$.}
\label{figure:phi}
\end{figure}

This leads to
\begin{align}
\label{asymptotic_representation}
n^{2/7}\int K_h(t-y)\,d\bigl(\hat G_n-G_0\bigr)(y)\sim n^{2/7}\int \th_{t,h,G_0}(e,s,\d)\left(\P_n-P_0\right)(e,s,\d),
\end{align}
where $\th_{t,h,G_0}$ is defined by (\ref{score}), where $G=G_0$, the underlying distribution function of the incubation time, and $\f$ is the solution of the equation (\ref{phi-eq}) and satisfies $\f(M_1)=0$. Moreover, (\ref{asymptotic_representation}) would imply:
\begin{align}
\label{asymptotic_normality}
n^{2/7}\int K_{h_n}(t-y)\,d\bigl(\hat G_n-G_0\bigr)(y)\stackrel{{\cal D}}\longrightarrow N(0,\s^2),
\end{align}
where
\begin{align*}
\s^2 = \lim_{n\to\infty}\text{var}\left(n^{2/7}\int \th_{t,h_n,G_0}(e,s,\d)\,d\P_n(e,s,\d)\right),
\end{align*}
and $n^{1/7}h_n\to c>0$. A picture of the function $\f$, solving (\ref{phi-eq}), is shown in Figure \ref{figure:phi}. This can be found by a simple iteration procedure for the integral equation (\ref{phi-eq2}) or a matrix equation after discretization, which can also be found in \cite{github:20}.

Note that, letting $\Phi(s)=\int_0^s\f(u)\,du$, and defining $0/0=0$.
\begin{align*}
&E\,\th_{t,h,G}(E,S,\dd)\\
&=\frac1{M}\int_{s\le e}\frac1e\frac{\f(s)}{G(s)}G(s)\,de\,ds
+\frac1{M}\int_{e<s}\frac1e\left\{\frac{\f(s)-\f(s-e)}{G(s)-G(s-e)}\right\}
\{G(s)-G(s-e)\}\,de\,ds\\
&=\frac1{M}\int_{s\le e}\frac1e\f(s)\,de\,ds+\frac1{M}\int_{e<s}\frac1e\{\f(s)-\f(s-e)\}\,de\,ds\\
&=\frac1{M}\int_{e=0}^M\frac1e\Phi(e)\,de+\frac1{M}\int_{e=0}^M\frac1e\{\Phi(M_1)-\Phi(e)-\Phi(M_1)+\Phi(0)\}\,de\\
&=\frac1{M}\int_{e=0}^M\frac1e\Phi(e)\,de-\frac1{M}\int_{e=0}^M\frac1e\Phi(e)\,de=0,
\end{align*}
using $\phi(s)=0$, $s\ge M_1$, were $M_1$ is the upper bound of the support of the density of the incubation time. Note that we use $M\ge M_1$, where $[0,M]$ is the interval containing the exit times (assumed to be uniformly distributed on $[0,M]$ in the simulation experiment).  In Figure \ref{figure:phi} we have $M_1=20$ and $M=30$. 
For the asymptotic variance, we get:
\begin{align}
\label{asymp_var}
&E\,\th_{t,h,G}(E,S,\dd)^2\nonumber\\
&=\frac1{M}\int_{s\le e}\frac1e\frac{\f(s)^2}{G(s)^2}G(s)\,de\,ds
+\frac1{M}\int_{s>e}\frac1e\left\{\frac{\f(s)-\f(s-e)}{G(s)-G(s-e)}\right\}^2
\{G(s)-G(s-e)\}\,de\,ds\nonumber\\
&=\frac1{M}\int_{s\le e}\frac{\f(s)^2}{e\,G(s)}\,de\,ds
+\frac1{M}\int_{s>e}\frac{\{\f(s)-\f(s-e)\}^2}{e\{G(s)-G(s-e)\}}\,de\,ds.
\end{align}
Note that:
\begin{align*}
&\text{var}\left(n^{2/7}\int K_h(t-y)\,d\hat G_n(y)\right)
\sim \text{var}\left(n^{2/7}\int \th_{t,h,G_0}(e,s,\d)\,d\P_n(e,s,\d)\right)\\
&=n^{-3/7}E\,\th_{t,h,G}(E,S,\dd)^2.
\end{align*}

\begin{figure}[!ht]
\centering
\includegraphics[width=0.5\textwidth]{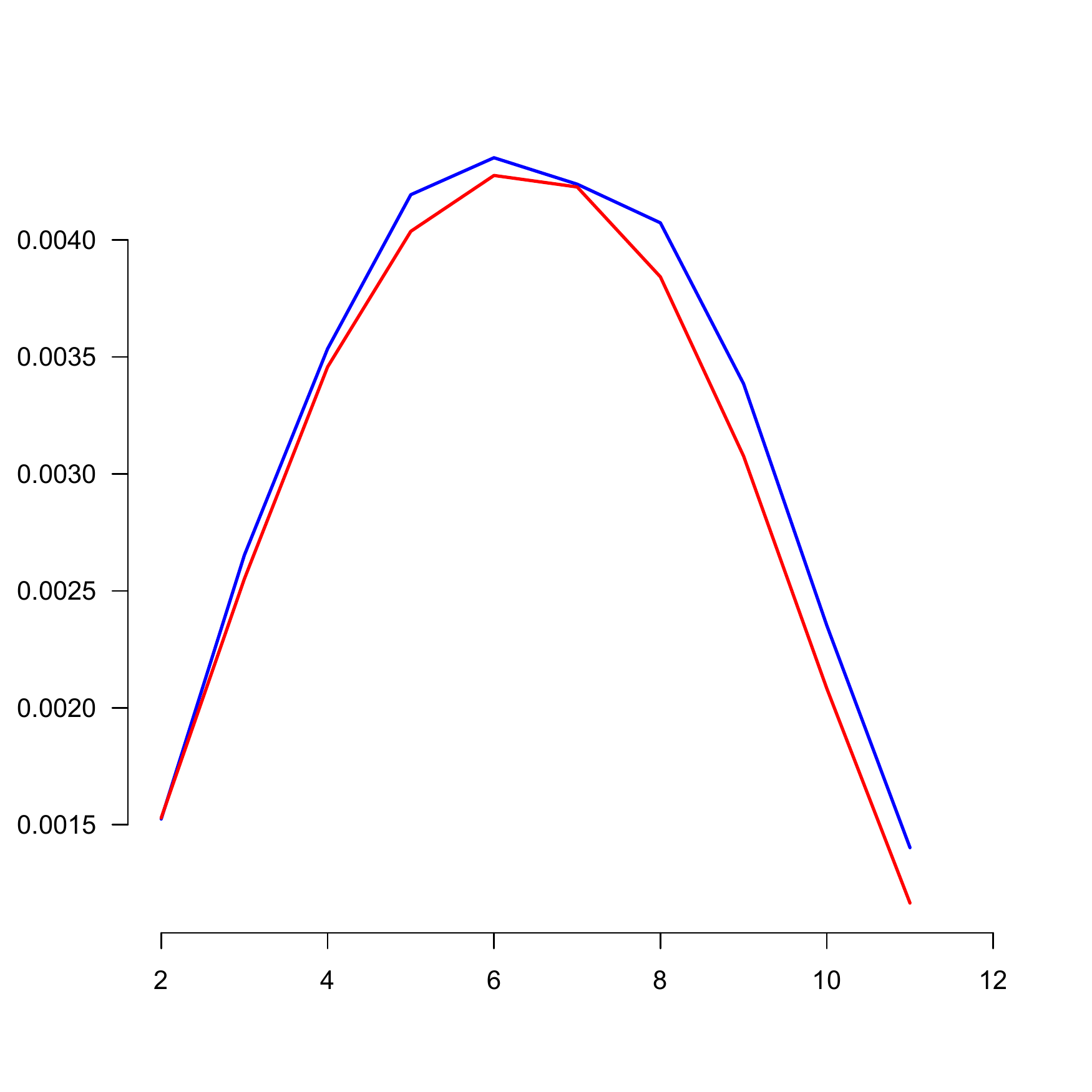}
\caption{Plot of Table \ref{tab:variances}. The variances in the simulation are given by the blue curve and the red curve gives the values $n^{-3/7}E\,\th_{t,h,G}(E,S,\dd)^2$, for $t=2,3,\dots,11$, $h=3.4$, where $G$ is the truncated Weibull distribution function.}
\label{figure:variances}
\end{figure}

A table for the variances of the density estimates at $t=2,3,\dots,11$, as computed from $1000$ samples of size $n=1000$ and from $n^{-3/7}E\,\th_{t,h,G}(E,S,\dd)^2$, as given by (\ref{asymp_var}).  The table is given graphically in Figure \ref{figure:variances}. 
\begin{table}
\begin{center}
\begin{tabular}{|c|c|c|}
\hline
$t$ &simulation variances &$\displaystyle{n^{-3/7}E\,\th_{t,h,G}(E,S,\dd)^2}$\\
\hline
2 	&0.001524376		&0.001528899\\
3 	&0.002652881		&0.002551415\\	
4 	&0.003535091		&0.003457335\\
5 	&0.004193696		&0.004037131\\
6 	&0.004351735		&0.004275926\\
7 	&0.004238654		&0.004226677\\
8 	&0.004073332		&0.003842444\\
9 	&0.003385165		&0.003076003\\
10  	&0.002352065		&0.002082613\\	
11  	&0.001402003		&0.001165108\\
\hline
\end{tabular}
\end{center}
\caption{A comparison of variances, given by a simulation of $1000$ samples of size $n=1000$ and the right-hand side of (\ref{asymp_var}). The bandwidth $h=3.4$ and $G$ is the distribution function of the Weibull distribution, truncated on the interval $[0,20]$.}
\label{tab:variances}
\end{table}

\bibliographystyle{apalike}
\bibliography{cupbook}

\begin{thebibliography}{}

\bibitem[Backer et~al., 2020]{backer:20}
Backer, J.~A., Klinkenberg, D., and Wallinga, J. (2020).
\newblock {I}ncubation period of 2019 novel coronavirus (2019-n{C}ov)
  infections among travellers from {W}uhan, {C}hina, 20-28 january 2020.
\newblock {\em Euro Surveill.}, 25.

\bibitem[Britton and Scalia~Tomba, 2019]{tom_gianpi:19}
Britton, T. and Scalia~Tomba, G. (2019).
\newblock Estimation in emerging epidemics: bases and remedies.
\newblock {\em J. R. Soc. Interface}, 16.

\bibitem[Groeneboom, 2020a]{github:20}
Groeneboom, P. (2020a).
\newblock Incubationtime.
\newblock \url{https://github.com/pietg/incubationtime}.

\bibitem[Groeneboom, 2020b]{pietNAW:20}
Groeneboom, P. (2020b).
\newblock {T}he {N}etherlands in {T}imes of {C}orona (in {D}utch).
\newblock {\em {N}ieuw {A}rchief voor {W}iskunde}, 21:181--184.

\bibitem[Groeneboom and Hendrickx, 2017a]{curstatCI:17}
Groeneboom, P. and Hendrickx, K. (2017a).
\newblock curstat{CI}.
\newblock R package.
\newblock Version 0.1.1.

\bibitem[Groeneboom and Hendrickx, 2017b]{kim_piet:17EJS}
Groeneboom, P. and Hendrickx, K. (2017b).
\newblock The nonparametric bootstrap for the current status model.
\newblock {\em Electron. J. Stat.}, 11(2):3446--3484.

\bibitem[Groeneboom and Jongbloed, 2014]{piet_geurt:14}
Groeneboom, P. and Jongbloed, G. (2014).
\newblock {\em Nonparametric Estimation under Shape Constraints}.
\newblock Cambridge Univ. Press, Cambridge.

\bibitem[Hall, 1990]{hall:90}
Hall, P. (1990).
\newblock Using the bootstrap to estimate mean squared error and select
  smoothing parameter in nonparametric problems.
\newblock {\em J. Multivariate Anal.}, 32:177--203.

\bibitem[Kolda et~al., 2003]{torczon2003}
Kolda, T.~G., Lewis, R.~M., and Torczon, V. (2003).
\newblock Optimization by direct search: new perspectives on some classical and
  modern methods.
\newblock {\em SIAM Rev.}, 45(3):385--482.

\bibitem[Reich et~al., 2009]{reich:09}
Reich, N.~G., Lessler, J., Cummings, D. A.~T., and Brookmeyer, R. (2009).
\newblock Estimating incubation period distributions with coarse data.
\newblock {\em Stat. Med.}, 28(22):2769--2784.

\bibitem[Sen and Xu, 2015]{SenXu2015}
Sen, B. and Xu, G. (2015).
\newblock Model based bootstrap methods for interval censored data.
\newblock {\em Comput. Statist. Data Anal.}, 81:121--129.

\bibitem[Silverman, 1986]{silverman:86}
Silverman, B. (1986).
\newblock {\em Density estimation for statistics and data analysis}, volume~26.
\newblock CRC press.

\bibitem[Torczon, 1997]{torczon:97}
Torczon, V. (1997).
\newblock On the convergence of pattern search algorithms.
\newblock {\em SIAM J. Optim.}, 7(1):1--25.

\end{thebibliography}

\end{document}